\documentclass[aps,prd,twocolumn,showpacs,amsmath,nofootinbib]{revtex4-1} 

\usepackage{graphicx}
\usepackage{epsfig}
\usepackage{color}

\usepackage{graphicx}
\usepackage{bm}
\usepackage{booktabs}
\usepackage{array}

\newcommand{\be}{\begin{equation}}
\newcommand{\ee}{\end{equation}}
\newcommand{\beq}{\begin{eqnarray}}
\newcommand{\eeq}{\end{eqnarray}}
\newcommand{\bef}{\begin{figure*}}
\newcommand{\eef}{\end{figure*}}

\begin{document}

\title{Constraints on the basic parameters of dark matter using the Planck data}

\author{Yupeng Yang$^{1,2}$}
\affiliation{$^1$Collage of Physics and Electrical Engineering, Anyang Normal University, 
Anyang, 455000, People's Republic of China\\
$^2$Joint Center for Particle, Nuclear Physics and Cosmology, Nanjing, 210093, People's Republic of China}

\begin{abstract}
Dark Matter annihilation or decay can affect the anisotropy of the cosmic 
microwave background (CMB). Therefore, the CMB data can be used to constrain the 
properties of a dark matter particle. In this work, we use the new CMB data obtained by the 
Planck satellite to investigate the limits on the 
basic parameters of a dark matter particle. The parameters are the dark matter mass ($m_{\chi}$) and 
the thermally averaged cross section ($\langle\sigma v\rangle$) for 
dark matter annihilation and the decay rate ($\Gamma$) (or lifetime $\tau = 1/\Gamma$) 
for dark matter decay. For dark matter annihilation we also consider 
the impact of the structure formation process which is neglected 
by the recent work. 
We find that for DM annihilation, the constraints on the parameters are 
$f_{ann}=\langle \sigma v\rangle /m_{\chi}< 0.16 \times 10^{-26} \mathrm{cm^{3}s^{-1}GeV^{-1}}$(or 
$f_{ann}<0.89 \times 10^{-6} \mathrm{m^{3}s^{-1}kg^{-1}}$, $95\%$ C.L.). For DM 
decay, the constraints on the decay rate are $\Gamma < 0.28 \times 10^{-25} \mathrm{s^{-1}}$($95\%$ C.L.).
\end{abstract}

\pacs{}

 \maketitle
\section{Introduction }
Dark matter (DM), the main components of the Universe, confirmed 
by many observations still keeps its nature mysterious~\cite{Jungman:1995df,Bertone2005279,bergstrom}. 
Its most widely 
accepted model, the weakly interacting massive particles (WIMPs) 
model, claims DM particles can self-annihilate into 
standard model particles, such as photons, electrons, and positrons~\cite{Jungman:1995df}, 
which might be observed by various experiments, such as 
PAMALA~\cite{Adriani:2008zr} and AMS-2~\cite{ams-2}. During the evolution of the 
Universe, 
these particles produced by DM annihilation interact with 
the medium of the Universe~\cite{Chen:2003gz,zhangle_ann,zhangle_dec}. 
For example, the photons produced through 
DM annihilation can ionize the hydrogens formed in the 
epoch of the recombination before appearance of the first stars. The changes 
of the ionization will be reflected in the anisotropy of the cosmic 
microwave background (CMB)~\cite{zhangle_ann,zhangle_dec,cmb_cons_slatyer_1,prd,dm_cons}. Therefore, the observation data of CMB can be 
used to investigate the nature of DM particles. Recently, the authors of \cite{dm_cons} used the 
Planck data to investigate the limits on the DM parameters, the mass $m_{\chi}$, 
and the thermally averaged cross section $\langle \sigma v \rangle$.  In that paper, 
they 
considered the uniform distribution of DM while neglecting 
the structure formation process of the Universe which claimed that the DM halos were formed 
in the redshift $z \sim 100$. Moreover, the subhalos or sub-subhalos are 
also formed in DM halos~\cite{Diemand:2005vz}. The DM annihilation rate can be enhanced 
in all these DM halos. In this work, we consider these effects. 
In addition to annihilation, DM particles are not stable in some models, and 
they can decay into the standard model particles~\cite{Takayama:2000uz,Arvanitaki:2008hq}. 
In this paper, we also 
use the Planck data to get the constraints on the decay rate (or lifetime) 
of DM. Not setting any specific DM model, the results of this work can be 
applied widely to those DM particles can annihilate or decay. 

This paper is organized as follows: In Sec. II, 
we briefly show how the DM particles' annihilation 
and decay affect the evolution of the Universe and we give the constraints on
the basic parameters of DM by use of the Planck data. 
We conclude in Sec. III.

\section{The impact of DM on the evolution of the Universe and 
constraints on the basic parameters}
DM particles can affect the Universe through 
the interaction between the medium of the Universe and the productions of DM annihilation or decay. 
The two main effects are to ionize the hydrogens and heat the medium~\cite{Chen:2003gz}. 
The changes of ionization with the redshift are governed by the equation~\cite{Chen:2003gz,zhangle_ann}

\be
(1+z)\frac{dx_{e}}{dz}=\left[R_{s}(z)-I_{s}(z)-I_{\chi}(z)\right]
\label{eq1}
\ee
where $I_{\chi}(z)$ is the ionization rate due to the DM, and $R_{s}(z)$ and 
$I_{s}(z)$ are the standard recombination rate and ionization rate by 
the standard sources, respectively. The ionization rate by the DM can be 
be written as 

\be
I_{\chi}(z)=\chi_{i}f(z)\Gamma_{ann} \frac{m_{\chi}}{{n_{b} E_{b}}}
\label{eq2}
\ee
for the DM annihilation and 

\be
I_{\chi}(z)=\chi_{i}f(z)\Gamma_{dec} \frac{\Omega_{\chi}}{\Omega_{b}}\frac{m_{b}}{n_{b}E_{b}}
\label{eq3}
\ee
for the DM decay. $\Gamma_{ann}$ and $\Gamma_{dec}$ 
(or $\Gamma_{dec}=1/\tau_{dec}$; $\tau_{dec}$ is the lifetime ) are the DM 
annihilation and decay rate, $n_b$ is the number density of baryon, 
and $E_{b} = 13.6 \mathrm{eV}$ is the ionization energy. 

If the structure formation effect is included, 
the DM annihilation rate can be written as 

\beq
\Gamma_{ann} = &&\frac{\langle\sigma v\rangle}{m_{\chi}}\rho_{cri}^2\Omega_{c}^2(1+z)^6B(z) \nonumber \\
=&&f_{ann}\rho_{cri}^2\Omega_{c}^2(1+z)^6B(z)
\eeq
where $B(z)$ is the 'boost factor' (or 'clumping factor') 
due to the structure formation effect; 
for more detailed discussions, one can see Ref.~\cite{boostfactor}. 
$\rho_{cri}$ is the critical density of the Universe. $f_{ann}$ is a new parameter, 
which is the combination of the basic parameters of DM, 
$f_{ann} = \langle\sigma v\rangle/m_{\chi}$.

For the boost factor $B(z)$, we followed the Ref.\cite{boostfactor} and used the 
form as

\beq
B(z) = 1+\frac{(1+z)^3}{\bar \rho^2_{DM}(z)}\int dM\frac{dn}{dM}(M,z)
\int \rho^2(r)dV
\eeq
where $dn/dM$ is the mass function of DM halos and we use the 
Press-Schechter formalism for our calculations~\cite{ps}. For the DM halos, we 
use the NFW (Navarro-Frenk-White) density profile. It is also found that there are many 
sub- and sub-substructures in DM halos~\cite{Diemand:2005vz}. The DM annihilation rate can 
be enhanced in these subhalos. In this paper, we include these subhalos while 
neglecting the contributions from the sub-subhalos. We set the smallest 
mass of DM halo as $\sim 10^{-6}M_\odot$~\cite{subhalo_mass}. We adopt that $\sim 10\%$ mass of DM 
halos is in the form of subhalos. We use the power law form of the 
mass function $\sim M^{1.95}$ for the subhalos~\cite{Diemand:2005vz}. 
The total boost factor of DM halos including the subhalos is~\cite{boostfactor} 

\be
B_{total} = 1+(B_{halos}-1)+(B_{subhalos}-1)
\ee

In equations (\ref{eq2}) and (\ref{eq3}), $\chi_{i}$ stands for 
the fraction of the energy which 
contributes to the ionization and is discussed first in Ref.~\cite{fraction_ion}. 
Here we adopt the form given by Ref.~\cite{Chen:2003gz}, 
$\chi_{i} = (1-x_{e})/3$. 
It should be noticed that although this form 
has been used frequently in previous works
~\cite{Chen:2003gz,zhangle_ann,zhangle_dec,prd}, it is not accurate. 
In the Ref.~\cite{ion_frac}, the difference of 
the constraints on DM parameters occured both in the approximate and the accurate 
have been investigated and only slightly differences 
of the upper limits for the present null detection of DM found. The similar 
discussions are also present in the Ref.~\cite{ion_frac_1}. 
$f(z)$ is the fraction of the energy which deposits 
to the medium of the Universe. It is different for different 
annihilation or decay channels and is a function of the redshift. 
Here we treat it as a free parameter. 
For more detailed discussions one can see 
Ref.~\cite{dm_cons}. In addition to the ionization, another important effect 
of DM to the evolution of the Universe is to heat the medium, e.g, the baryonic 
gas. The evolution of the gas temperature can be written as

\beq
(1+z)\frac{dT_{b}}{dz}&& =\frac{8\sigma_{T}a_{R}T^4_{cmb}}{3m_{e}cH(z)}
\frac{x_e}{1+f_{He}+x_e}(T_{b}-T_{cmb}) \nonumber \\ 
&&-\frac{2}{3k_{B}H(z)}\frac{K_{\chi}}
{1+f_{He}+x_{e}}+2T_{b} 
\eeq
where $K_{\chi}$ is similar to $I_{\chi}$ and stands for the heating rate 
to the medium by the DM. Following Ref.~\cite{Chen:2003gz}, we use the form as 
$K_{\chi} = (1+2x_{e})/3$.

Recently, the authors of ~\cite{dm_cons} used the Planck data to get the constraints 
on the DM parameters for the annihilation case, but they did not consider the 
effect of the structure formation. 
In this work, we include this effect, and we also get the constraints 
on the decay rate (or lifetime) for the DM decay. Because the limits are independent 
on any specific DM models, the results can be used widely for many DM annihilation or decay 
models. 

We modified the public code RECFAST~\footnote{http://camb.info}to included the DM effect, and we used 
the public code CosmoMC~\footnote{http://cosmologist.info/cosmomc/} 
to get the constraints on the parameters. 
We consider six cosmological parameters and a new parameter 

\be
\{\Omega_{b}h^2,\Omega_{c}h^2, \theta, \tau, n_s, A_s, F_{ann}\}, 
\ee
for the DM annihilation and 
\be
\{\Omega_{b}h^2,\Omega_{c}h^2, \theta, \tau, n_s, A_s, F_{dec}\}, 
\ee
for the DM decay. Here are the new parameters: 
$F_{ann} = f(z)f_{ann}=f(z)\langle\sigma v\rangle /m_{\chi}$ 
and $F_{dec} = \Gamma_{dec}f(z)$. 

The final constraints on the parameters are given in Table~\ref{tab_ann} 
for the DM annihilation and Table~\ref{tab_dec} for the DM decay. 
For making comparisons, the results for the 
case of smooth DM distribution are given too. In this case, 
the 'boost factor' is $B(z)=1$.
From these results, it can be seen that for $f(z) =1 $, the $95\%$ upper limits are 

\be
\frac{\langle\sigma v\rangle}{m_{\chi}} < 0.16(0.24) \times 10^{-26} \mathrm{cm^{3}s^{-1}GeV^{-1}}
\ee

for the DM annihilation case (0.24 is the case of smooth distribution) and 

\be
\Gamma < 0.28 \times 10^{-25} \mathrm{s^{-1}}
\ee

for the DM decay case. 

For the DM annihilation case, our results are consistent with Ref.~\cite{dm_cons}(e.g., Table II). 
For this point, we can convert our results as 

\beq
\frac{\langle\sigma v\rangle}{m_{\chi}} &&< 0.16(0.24) \times 10^{-26} \mathrm{cm^{3}s^{-1}GeV^{-1}} \nonumber \\
&&=0.89(1.34) \times 10^{-6} \mathrm{m^{3}s^{-1}kg^{-1}}
\eeq

We also plot the 1D and 2D probability distributions in Figs.~\ref{ann_1D} and 
\ref{ann_2D} for the DM annihilation (including the smooth distribution case, dotted lines) 
and Figs.~\ref{dec_1D} and \ref{dec_2D} 
for the DM decay, respectively. From these plots, one can find that the correlation 
between the $F_{ann}$ parameter and the cosmological parameters is stronger 
for the DM annihilation than that of $F_{dec}$ for the DM decay. The main 
reason is that the DM annihilation rate is proportional to the number density 
square, so the effects due to the annihilation are very strong during the 
recombination. For the DM annihilation, the differences between 
the clumpy and smooth DM distribution cases are not so huge. These results 
indicate once again that the limits on the DM parameters are mainly from the epoch of recombination.

\begin{table*}[h]
\caption{Posterior constraints on the parameters 
for the DM annihilation. The first and second lines of every item correspond to the 
clumpy and smooth DM distribution respectively.}
\label{tab_ann}
\begin{center}     
\begin{ruledtabular}  
\begin{tabular}{cccccccc}
Parameter & $\Omega_{b}h^2$ & $\Omega_{c}h^2$ & $100\theta$ &$\tau$&$n_s$&$ln(10^{10}A_s)$&$F_{ann}\mathrm{(10^{-26}cm^{3}s^{-1}GeV^{-1})}$\\
\noalign{\smallskip}\hline\noalign{\smallskip}
Mean     & 0.02207 & 0.1196 & 1.0412 &0.088&0.962&3.09&0.053\\
         & 0.02209 & 0.1195 & 1.0413 &0.089&0.963&3.10&0.070\\
$2\sigma$ low  &0.02155 &0.1147 &1.0401&0.061&0.948&3.05&0  \\
               &0.02152 &0.1140 &1.0401&0.062&0.949&3.05&0   \\
$2\sigma$ up &0.02260&0.1251&1.0425&0.1149&0.977&3.15&0.16 \\
             &0.02265&0.1244&1.0425&0.1154&0.978&3.15&0.24 \\
\end{tabular}
\end{ruledtabular}
\end{center}
\end{table*}

\begin{table*}[h]
\caption{Posterior constraints on the parameters 
for the DM decay.}
\label{tab_dec}   
\begin{center}   
\begin{ruledtabular}
\begin{tabular}{cccccccc}
Parameter & $\Omega_{b}h^2$ & $\Omega_{c}h^2$ & $100\theta$ &$\tau$&$n_s$&$ln(10^{10}A_s)$&$F_{dec} (10^{-25} s^{-1})$ \\
\noalign{\smallskip}\hline\noalign{\smallskip} 
Mean     &0.02205 & 0.1199 &1.0413 &0.088&0.960&3.09&0.077\\
$2\sigma$ low  &0.02152 &0.1147 &1.0401&0.064&0.94&3.04&0  \\
$2\sigma$ up &0.02262&0.1251&1.0425&0.115&0.97&3.14&0.28 \\
\end{tabular}
\end{ruledtabular}
\end{center}
\end{table*}

\section{summary and discussion}
In this work, we used the new data from the Planck satellite to investigate the 
limits on the DM basic parameters for the annihilation and decay. 
By considering the structure formation effect for the DM 
annihilation case, we found that the constraints on the $f_{ann}$ parameter 
are $f_{ann} < 0.16(0.24) \times 10^{-26} \mathrm{cm^{3}s^{-1}GeV^{-1}}$ or 
$f_{ann} < 0.89(1.34) \times 10^{-6} \mathrm{m^{3}s^{-1}kg^{-1}}$($95\%$ C.L.). For the DM decay, 
the constraints on the decay rate are $\Gamma < 0.28 \times 10^{-25} s^{-1}$($95\%$ C.L.).

As mentioned in the Sec.II, for the clumpy DM distribution, 
the smallest mass of DM halo is set as $\sim 10^{-6}M_{\odot}$, 
which is different for different DM models. In theory for WIMPs DM, 
this value ranges from $10^{-12}M_{\odot}$ to $10^{-4}M_{\odot}$ for typical kinetic decoupling 
temperatures. From the results of current numerical simulations, the typical smallest mass of DM halos 
is $\sim 10^{6}M_{\odot}$. In Ref.~\cite{boostfactor}, the authors discussed the effects on the 'boost factor' for the different values of the smallest 
DM halos. They found that there are differences of $\sim 2$ orders of magnitude 
of the 'boost factor' 
for DM halo mass $10^{-12}M_{\odot}$ and $10^{-4}M_{\odot}$ at $z \sim 50$ 
(upper panel of Fig. 1 in Ref.~\cite{boostfactor}). For DM halo mass $10^{6}M_{\odot}$, 
the differences of $\sim 5$ orders of magnitude differences are present 
at $z \sim 20$ compared with the DM smooth distribution. 
Therefore, the largest differences usually appear in nearby universe, 
and it is believed that the changes of limits on the DM parameters are slight 
if one change the values of the smallest mass of DM halos.

Another factor which can affect the limits on the DM is the 
density profile of DM halos. In this work, we have used the NFW profile,
which is well in fitting many observations data. 
In addition there are still many other observations or N-body simulations 
which are favored by the other profiles, 
such as Einasto profile~\cite{einasto_1,einasto_2,einasto_3,den_pro}, 
which are slightly different from that of NFW profile for the final constrains.

One point in this work that should be noticed is that we have set $f(z)$ as a free parameter, and 
for the final constraints we have set $f(z) =1$, which means that all the energy 
produced by the DM annihilation or decay has deposited into the medium of the 
Universe. In Ref.~\cite{dm_cons}, the dependence of $f(z)$ 
on the redshift and different annihilation channels were discussed by the authors. 
It can be seen that the final constraints are slightly 
different (Table II of Ref.~\cite{dm_cons}.

\bef
\epsfig{file=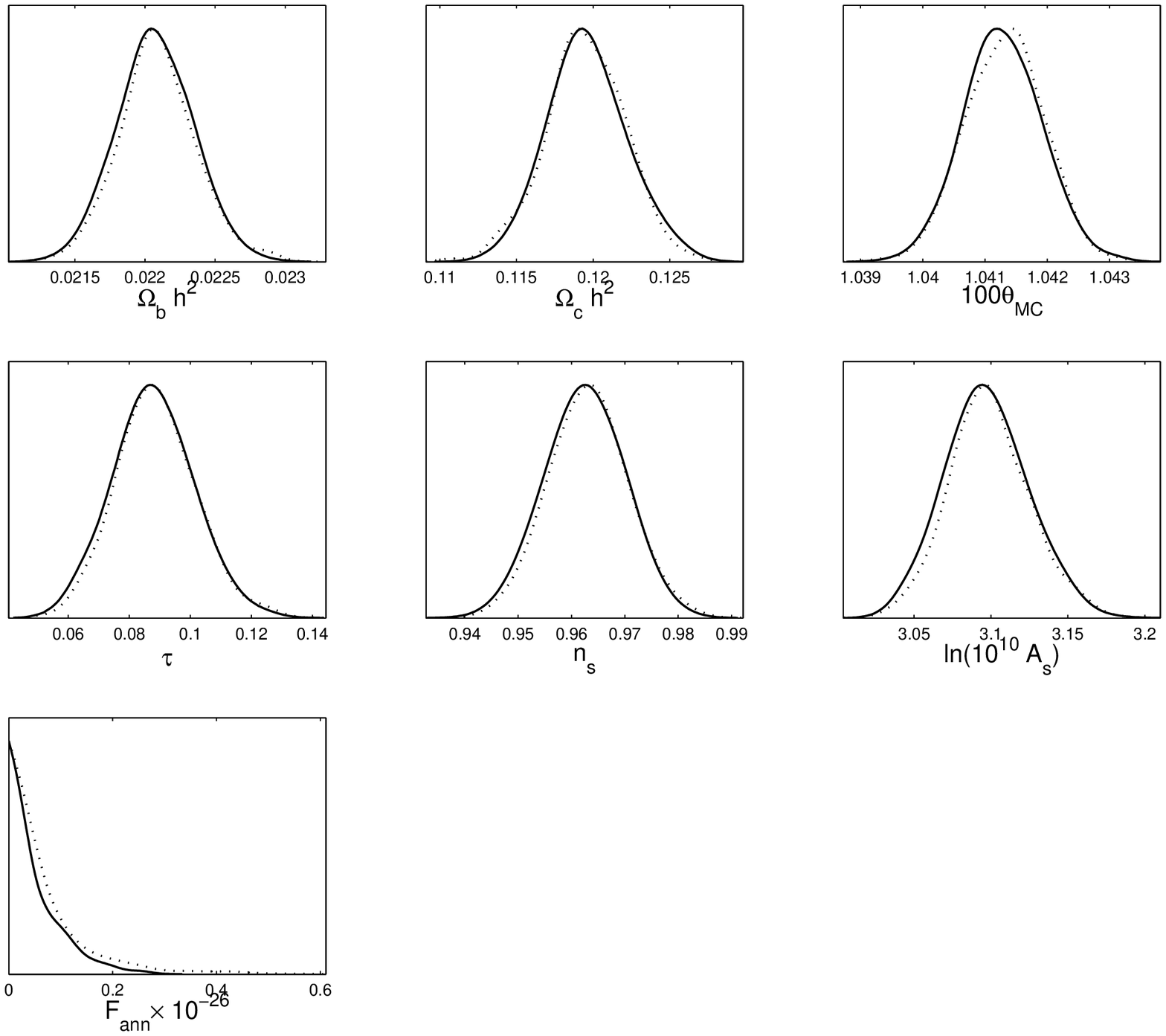,width=0.5\textwidth}
\caption{The marginalized probability distribution function of 
parameters for the DM annihilation case. The solid and dotted lines correspond 
to the clumpy and smooth DM distribution respectively.}
\label{ann_1D}
\eef

\bef
\epsfig{file=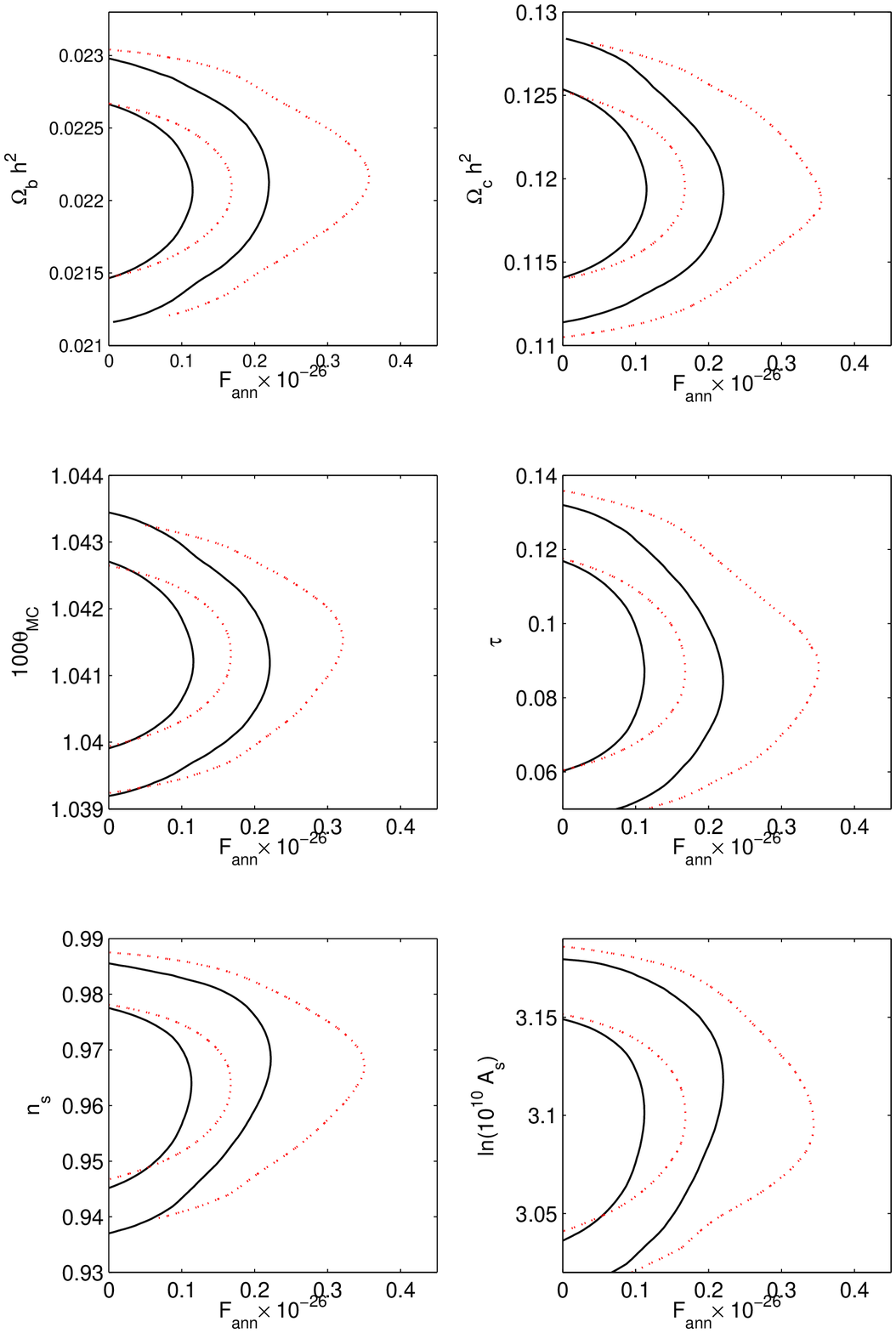,width=0.5\textwidth}
\caption{The 2D contours distribution function of 
parameters for the DM annihilation case($68\%$ and $95\%$ confidence level). The solid (black) and dotted (red) lines correspond 
to the clumpy and smooth DM distribution respectively.}
\label{ann_2D}
\eef

\bef
\epsfig{file=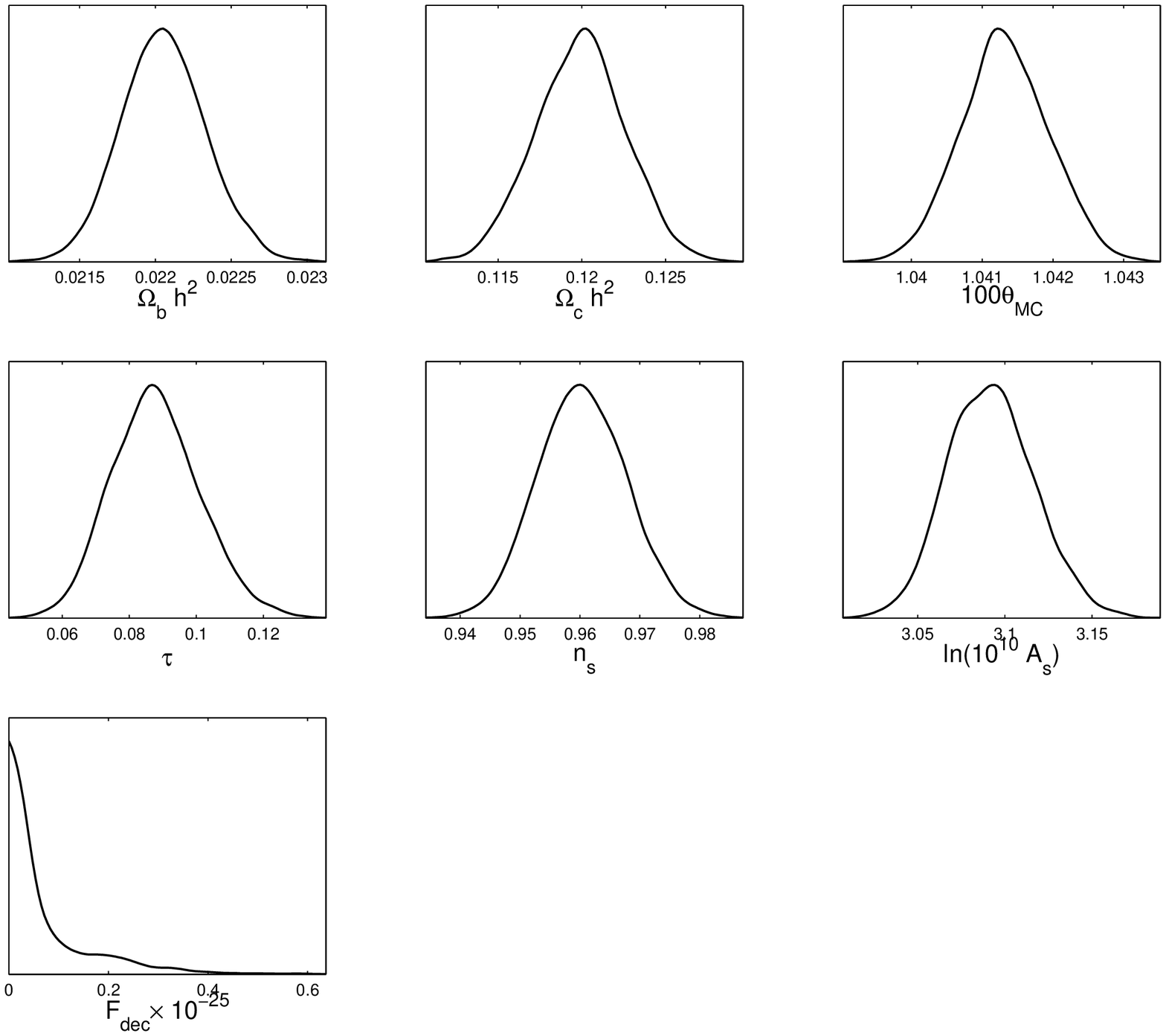,width=0.5\textwidth}
\caption{The marginalized probability distribution function of 
parameters for the DM decay case.}
\label{dec_1D}
\eef

\bef
\epsfig{file=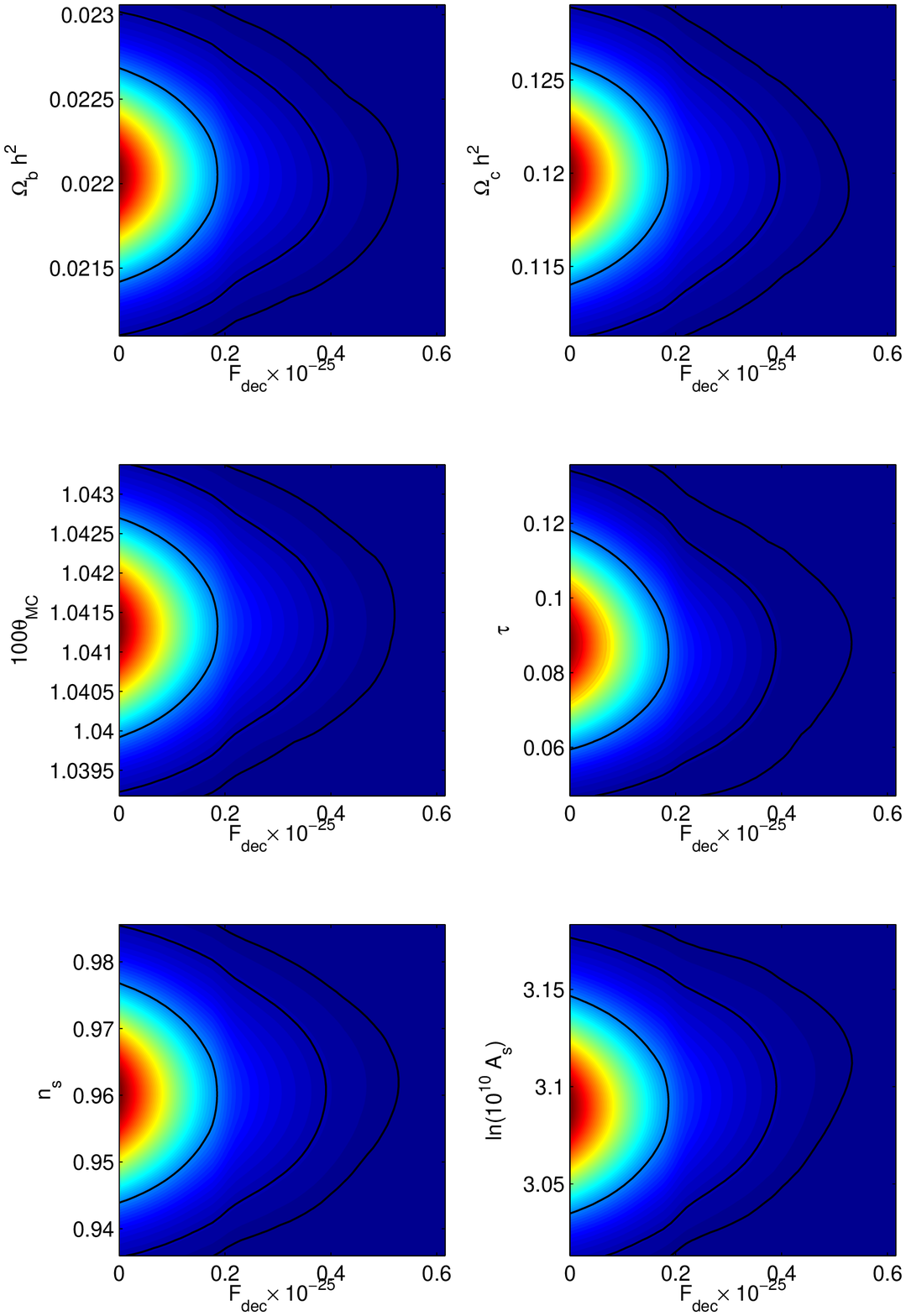,width=0.5\textwidth}
\caption{The 2D contours distribution function of 
parameters for the DM annihilation case($68\%$, $95\%$, and $99\%$ confidence level).}
\label{dec_2D}
\eef

\emph{NOTE: the latest constraints on the DM parameters for annihilation can 
be found in the paper of Planck Collaboration (arXiv:1502.01589).}

\section{Acknowledgments} The author wishes to thank Yichao Li for
the useful discussions on the CosmoMC for the Planck data and 
thanks Yan Qin for improving the manuscript. Y. Y. thanks Dr. Xiaoyuan Huang 
and Prof. Xuelei Chen. We also thank the referees for the very 
useful comments. Our MCMC chains computation was performed on the
SGC(ShuGuang Cluster) of the Anyang Normal University. Yupeng Yang 
is supported by the National Science Foundation of China (Grants
No.11347148, No.U1404114 and No.11373068).


\end{document}